\documentclass[pre,amsmath,amssymb,nofootinbib,floatfix,superscriptaddress]{revtex4}

\usepackage{graphicx,bm}
\usepackage[normalem]{ulem} 

\makeatletter
\def\graphicscale{\twocolumn@sw{0.3}{0.4}}
\def\graphicthreescale{\twocolumn@sw{0.3}{0.4}}

\begin{document}

\title{Comment on {\em Machine Learning the Operator Content of the
    Critical Self-Dual Ising-Higgs Gauge Model}, arXiv:2311.17994v1 }

\author{Claudio Bonati} \affiliation{Dipartimento di Fisica
  dell'Universit\`a di Pisa and INFN Largo Pontecorvo 3, I-56127 Pisa,
  Italy}

\author{Andrea Pelissetto}
\affiliation{Dipartimento di Fisica dell'Universit\`a di Roma Sapienza
        and INFN Sezione di Roma I, I-00185 Roma, Italy}

\author{Ettore Vicari} 
\affiliation{Dipartimento di Fisica dell'Universit\`a di Pisa,
        Largo Pontecorvo 3, I-56127 Pisa, Italy}

\date{\today}

\begin{abstract}
  We critically discuss the results reported in arXiv:2311.17994v1 by
  L. Oppenheim, M. Koch-Janusz, S. Gazit, and Z. Ringel, on the
  multicritical behavior of the three-dimensional Ising-Gauge model at
  the multicritical point. We argue that their results do not
  contradict the theoretical scenario put forward in {\em
    Multicritical point of the three-dimensional ${\mathbb Z}_2$ gauge
    Higgs model}, Phys. Rev. B \textbf{105}, 165138 (2022),
  arXiv:2112.01824, that predicted a multicritical behavior controlled
  by the stable $XY$ fixed point of an effective three-dimensional
  ${\mathbb Z}_2\oplus {\mathbb Z}_2$ Landau-Ginzburg-Wilson $\Phi^4$
  field theory. Actually, their results, as well as all numerical
  results reported so far in the literature, are consistent with a
  multicritical $XY$ scenario.
\end{abstract}

\maketitle

The three-dimensional (3D) Ising-Higgs, or ${\mathbb Z}_2$ gauge
Higgs, model considered in Ref.~\cite{OKGR-23} is a gauge-spin model
defined on a cubic lattice. The fundamental fields are ${\mathbb Z}_2$
(Ising) spins $s_{\bm x}=\pm 1$ and ${\mathbb Z}_2$ gauge variables
$\sigma_{{\bm x},\mu}=\pm 1$, associated with the lattices sites and
bonds, respectively.  The Hamiltonian
reads~\cite{Wegner-71,BDI-75,FS-79,Kogut-79}
\begin{eqnarray}
  H = - J \sum_{{\bm x},\mu} s_{\bm x} \, \sigma_{{\bm x},\mu} \,
  s_{{\bm x}+\hat{\mu}} - \kappa \sum_{{\bm x},\mu>\nu}
  \sigma_{{\bm x},\mu} \,\sigma_{{\bm
    x}+\hat{\mu},\nu} \,\sigma_{{\bm x}+\hat{\nu},\mu} \,\sigma_{{\bm
    x},\nu}.
\label{HiggsH}
\end{eqnarray}

\begin{figure}[tbp]
\includegraphics[width=0.5\columnwidth, clip]{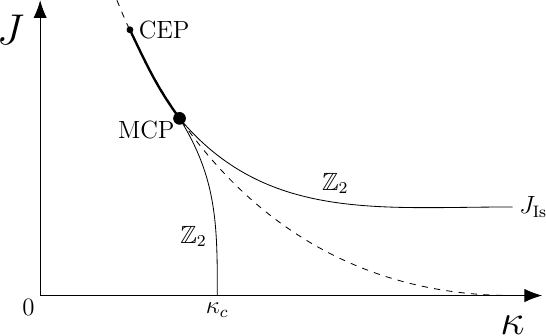}
\caption{Sketch of the phase diagram of the 3D ${\mathbb Z}_2$ gauge
  Higgs model (\ref{HiggsH}). The dashed and thick lines correspond to 
  the self-dual line. Along the thick line first-order transitions occur.
  The two lines labeled ``${\mathbb Z}_2$" are related by duality and
  correspond to Ising transitions.  The three lines meet at a MCP 
  on the self-dual line.}
\label{phadia}
\end{figure}

The studies reported in Refs.~\cite{OKGR-23,BPV-22} focused on the behavior at
the multicritical point (MCP) located along the self-dual line, see
Fig.~\ref{phadia}.  Refs.~\cite{OKGR-23,BPV-22,SSN-21} determined the critical
exponents at the MCP, obtaining results that are in very good agreement with
each other, and that apparently indicate that the MCP is an $XY$ multicritical
point (the numerical estimates of the critical exponents differ from the most
accurate estimates of the $XY$ exponents by approximately 1\%).  This
multicritical $XY$ scenario would imply that the critical correlations at the
MCP experience an effective enlarged O(2) global symmetry.  To verify this
possibility, Ref.~\cite{OKGR-23} investigated if there exists a conserved
current, i.e., a vector operator of renormalization-group (RG) dimension 2,
associated with the emerging global continuous O(2) symmetry.  They used a
sophisticated numerical method, looking for the putative conserved current in
the space of essentially local and relatively simple operators defined on small
lattices (in the language appropriate for 2+1 systems, their operators are
local both in space and in the imaginary-time direction). They did not find any
such operator.

We would like to stress that this result is {\em not} in contradiction with the
interpretation of the numerical results proposed in Ref.~\cite{BPV-22}. The
multicritical transition is due to the interplay of gauge and spin degrees of
freedom. The first (rather standard) assumption made in Ref.~\cite{BPV-22} was
that one can obtain an effective description by considering the order
parameters that characterize the critical behavior along the two lines that
meet at the MCP (the ${\mathbb Z}_2$ lines in the figure). On one side, the
transition is of magnetic type and thus we expect the order parameter to be a
scalar field $\phi_1$. The gauge transition is a topological transition and
therefore it is characterized by the absence of a gauge-invariant local order
parameter. The order parameter should be a nonlocal function of the gauge
degrees of freedom. Now, it comes the second, more delicate assumption, i.e.
that one can extend the duality, which is formally defined as a mapping of the
Hamiltonian parameters that essentially preserves the free energy, to a mapping
of fields and operators.  Using this mapping, we argue, we can redefine the
fields so that the nonlocal gauge order parameter is mapped into a scalar field
$\phi_2$ that interacts locally with $\phi_1$. If this conjecture holds, one
can explain the observed multicritical $XY$ behavior in a standard fashion. The
multicritical behavior is effectively described by the multicritical ${\mathbb
Z}_2\oplus {\mathbb Z}_2$ Landau-Ginzburg-Wilson (LGW) field theory, with
Hamiltonian density
\begin{eqnarray}
{\cal H} = \frac{1}{2} \Bigl[ ( \partial_\mu \varphi_1)^2 + (
  \partial_\mu \varphi_2)^2\Bigr] + \frac{1}{2} \Bigl( r_1 \varphi_1^2
+ r_2 \varphi_2^2 \Bigr) + \frac{1}{4!} \Bigl[ v_1
  \varphi_1^4 + v_2 \varphi_2^4 + 2 w\, \varphi_1^2\varphi_2^2 \Bigr].
\label{bicrHH} 
\end{eqnarray}
This model has been extensively
studied~\cite{FN-74,NKF-74,PV-02,CPV-03,HV-11,BPV-22}.  In particular, it
admits a bicritical point with a critical behavior controlled by a stable $XY$
fixed point.  The main conjecture of Ref.~\cite{BPV-22} is that this stable
$XY$ fixed point is the relevant one that controls the multicritical $XY$
behavior in the $\mathbb{Z}_2$ Higgs theory.  Obviously, in the LGW approach
there exists a local current that is conserved at the $XY$ fixed point and that
therefore characterizes the O(2) enlargement of the symmetry. However, once we
go back to the original gauge variables by the conjectured duality mapping, the
current should become a nonlocal function of the gauge variables.  Thus, the
nonexistence of a {\em local} conserved current within the lattice
$\mathbb{Z}_2$ Higgs theory does not contradict the conjecture of
Ref.~\cite{BPV-22}.  Note that the expected behavior at the MCP differs
substantially from that observed in the 3D lattice Ashkin-Teller model, also
considered in Ref.~\cite{OKGR-23}, in which there is a standard and
straightforward method to identify the correct conserved current.

Finally, we point out that one should distinguish the standard $XY$ point of an
O(2) vector model from that emerging at a ${\mathbb Z}_2\oplus {\mathbb Z}_2$
MCP. In the RG language, the relevant multicritical fixed point is obtained by
performing RG transformations in a different class of Hamiltonians (those with
${\mathbb Z}_2\oplus {\mathbb Z}_2$ symmetry) and therefore it cannot be
exactly identified with the standard $XY$ one. For instance, at the MCP there
are additional relevant and irrelevant operators that break the O(2)
symmetry~\cite{BPV-22}.

In conclusion, on the basis of the above arguments, we believe that the final
claim of Ref.~\cite{OKGR-23}, i.e. that their results are ``strikingly
inconsistent'' with the scenario outlined in Ref.~\cite{BPV-22}, is incorrect.
Actually, they provide further support to the multicritical $XY$ scenario put
forward in Ref.~\cite{BPV-22}.

\end{document}